\documentclass[aps,pra,groupedaddress,nofootinbib,notitlepage,showpacs,floatfix,superscriptaddress]{revtex4-1}
\usepackage{color,graphicx,graphics,times,bm,amssymb,amsmath,amsfonts,amsthm,mathrsfs,xr,lineno,soul}
\soulregister\cite7
\soulregister\ref7
\soulregister\eqref7

\DeclareMathOperator*{\argmin}{arg\,min}
\newcommand{\beq}{\begin{equation}}
\newcommand{\eeq}{\end{equation}}

\usepackage{hyperref}
\hypersetup{
    colorlinks=true,       
    linkcolor=cyan,          
    citecolor=magenta,        
    filecolor=magenta,      
    urlcolor=cyan,           
    runcolor=cyan
}

\renewcommand*{\thefootnote}{\fnsymbol{footnote}}
\makeatletter
\newcommand\footnoteref[1]{\protected@xdef\@thefnmark{\ref{#1}}\@footnotemark}
\makeatother

\newcommand\blfootnote[1]{%
  \begingroup
  \renewcommand\thefootnote{}\footnote{#1}%
  \addtocounter{footnote}{-1}%
  \endgroup
}

\begin{document}

\title{Quantum annealing versus classical machine learning applied to a simplified computational biology problem
} 
\author{Richard Li}
\affiliation{Department of Chemistry, University of Southern California, Los Angeles, California 90089, USA}
\affiliation{Computational Biology and Bioinformatics Program, Department of Biological Sciences, University of Southern California, Los Angeles, CA 90089, USA} 
\affiliation{Center for Quantum Information Science \& Technology, University of Southern California, Los Angeles, California 90089, USA}
\author{Rosa Di Felice}
\affiliation{Computational Biology and Bioinformatics Program, Department of Biological Sciences, University of Southern California, Los Angeles, CA 90089, USA} 
\affiliation{Department of Physics and Astronomy, University of Southern California, Los Angeles, California 90089, USA}
\affiliation{Center S3, CNR Institute of Nanoscience, Via Campi 213/A, 41125 Modena, Italy}
\author{Remo Rohs}\blfootnote{Correspondence and requests for materials should be addressed to R.R. (rohs@usc.edu) or D.A.L. (lidar@usc.edu)}
\affiliation{Department of Chemistry, University of Southern California, Los Angeles, California 90089, USA}
\affiliation{Computational Biology and Bioinformatics Program, Department of Biological Sciences, University of Southern California, Los Angeles, CA 90089, USA} 
\affiliation{Department of Physics and Astronomy, University of Southern California, Los Angeles, California 90089, USA}
\affiliation{Department of Computer Science, University of Southern California, Los Angeles, CA 90089, USA}
 \author{Daniel Lidar}
 \affiliation{Department of Chemistry, University of Southern California, Los Angeles, California 90089, USA}
 \affiliation{Center for Quantum Information Science \& Technology, University of Southern California, Los Angeles, California 90089, USA}
\affiliation{Department of Physics and Astronomy, University of Southern California, Los Angeles, California 90089, USA}
\affiliation{Department of Electrical Engineering, University of Southern California, Los Angeles, California 90089, USA}

\date{\today}
\begin{abstract}
Transcription factors regulate gene expression, but how these proteins recognize and specifically bind to their DNA targets is still debated. Machine learning models are effective means to reveal interaction mechanisms. Here we studied the ability of a quantum machine learning approach to predict binding specificity. Using simplified datasets of a small number of DNA sequences derived from actual binding affinity experiments, we trained a commercially available quantum annealer to classify and rank transcription factor binding. The results were compared to state-of-the-art classical approaches for the same simplified datasets, including simulated annealing, simulated quantum annealing, multiple linear regression, LASSO, and extreme gradient boosting. Despite technological limitations, we find a slight advantage in classification performance and nearly equal ranking performance using the quantum annealer for these fairly small training data sets. Thus, we propose that quantum annealing might be an effective method to implement machine learning for certain computational biology problems.
\end{abstract}

\maketitle

\section*{Introduction}
Quantum computation has been the subject of intense scientific scrutiny for its potential to solve certain fundamental problems, such as factoring of integers \cite{Shor:97} or simulation of quantum systems \cite{Feynman:1985ul,Lloyd:96}, more efficiently than classical algorithms, by using unique quantum phenomena including entanglement and tunneling. More recently, there has been much interest in the potential of quantum machine learning to outperform its classical counterparts \cite{Neven1,Pudenz:2013kx,Lloyd:2013aa,Wittek:2014aa,Lloyd:2014fk,Rebentrost:2014uq,Wiebe:2014eu,Schuld:2014eu,Wiebe:2014vn,Aaronson:2015lh,Adachi:2015qe,Amin:2016,Biamonte:2016aa,Benedetti:2016oz,Mott:2017aa}. Although different implementations and models of quantum computing are still in development, promising theoretical and experimental research indicates that quantum annealing  (QA) \cite{kadowaki_quantum_1998}, or adiabatic quantum optimization \cite{farhi_quantum_2001}, may be capable of providing advantages in solving classically-hard problems that are of practical interest (for a review see Ref.~\cite{Albash-Lidar:RMP}). QA is the only paradigm of quantum computation that currently offers physical implementations of a non-trivial size, namely the D-Wave processors \cite{Harris:2010kx,Dwave,Bunyk:2014hb}.

\begin{figure}[htp]
 \includegraphics[scale=0.8]{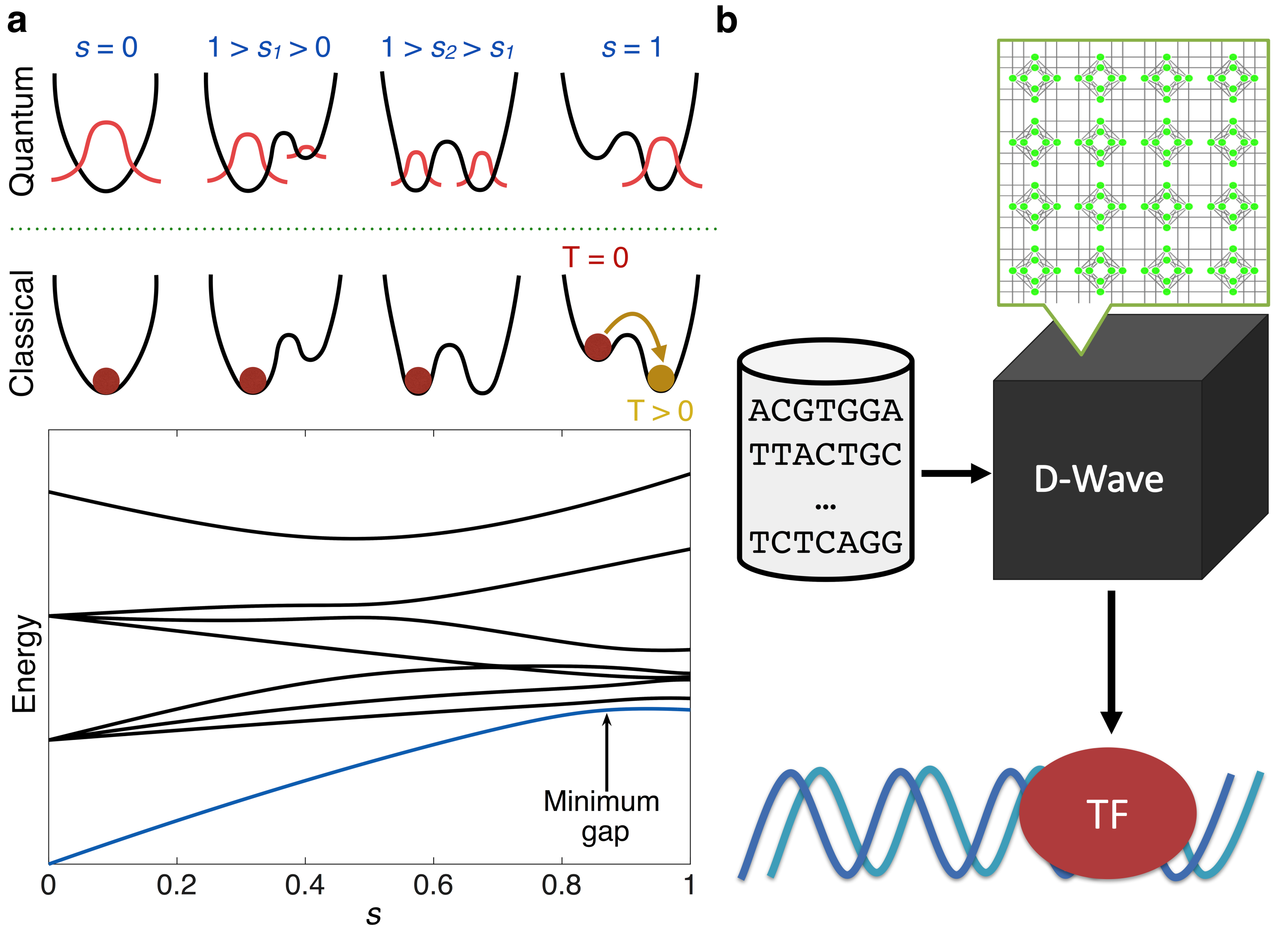}
\caption{Illustration of the principles and purpose of this work. (\textbf{a}) Top: Quantum versus classical adiabatic annealing processes. With quantum annealing there is the possibility for the system state (red) to tunnel through a changing barrier (black) and arrive at the ground state; for classical annealing, the system must rely on thermal fluctuations (temperature $T>0$) to overcome any energy barriers. Bottom: Typical spectrum of instantaneous energy eigenvalues during adiabatic quantum optimization. The ground state energy at $s=0$ has a significant gap to the next energy level. The speed at which the optimization can take place depends on the size of the minimum gap. (\textbf{b}) Using simplified datasets of a small number of sequences derived from actual binding affinity experiments, we use D-Wave, a commercially available quantum annealer, to classify and rank binding affinity preferences of a TF to DNA.}\label{fig:introfig}
\end{figure}

The adiabatic theorem of quantum mechanics, which underlies QA, implies that a physical system will remain in the ground state if a given perturbation acts slowly enough and if there is a gap between the ground state and the rest of the system's energy spectrum \cite{Kato:50} (Fig.~\ref{fig:introfig}a). To use the adiabatic theorem to solve optimization problems, we can specify an initial Hamiltonian, $H_B$, whose ground state is easy to find (typically a transverse field), and a problem Hamiltonian, $H_P$, that does not commute with $H_B$ and whose ground state encodes the solution to the problem we are seeking to optimize \cite{Farhi:00}. We then interpolate from $H_B$ to $H_P$ by defining the combined Hamiltonian $H(s) = A(s)H_B + B(s) H_P$, with $0\le s=t/t_f \le 1$, where $A(s)$ and $B(s)$ are, respectively, decreasing and increasing smoothly and monotonically, $t$ is time, and $t_f$ is the total evolution, or annealing time. The adiabatic theorem ensures that the ground state of the system at $t=t_f$ will give the desired solution to the problem, provided the interpolation is sufficiently slow, i.e., $t_f$ is large compared to the timescale set by the inverse of the smallest ground state gap of $H(s)$ and by $dH(s)/ds$ \cite{Jansen:07} (Fig.~\ref{fig:introfig}a). When QA is implemented in a physical device, temperature and other noise effects play an important role; thermal excitation and relaxation cannot be neglected and affect performance \cite{childs_robustness_2001,amin_decoherence_2009,Albash:2015nx}.

Quantum annealing algorithms were implemented on the D-Wave Two X (DW2X) processor installed at the Information Sciences Institute of the University of Southern California. The problem Hamiltonians that are used for D-Wave (DW) can be described as Ising spin models with tunable parameters \cite{Barahona1982}. The Ising model assumes a graph $G = (V,E)$ composed of a set of vertices, $V$, and edges, $E$. Each of the $N$ spins is a binary variable located at a unique vertex. For the DW2X, $N=1098$, the spins are represented by superconducting flux qubits, and $G$ is the so-called Chimera graph (see Supplementary Material, Fig.~S1). The problem, or Ising, Hamiltonian for this system can be written as
\beq
H_P = \sum_{i\in V} h_i\sigma_i^z + \sum_{(i,j)\in E}J_{ij}\sigma_i^z\sigma_j^z
\label{eq:ising},
\eeq
where the local fields $\{h_i\}$ and couplings $\{J_{ij}\}$ define a problem instance, and are programmable on the DW2X to within a few percent Gaussian distributed error. The $\sigma_i^z$ represent both binary variables taking on values $\pm 1$, and the Pauli $z$-matrices. Given a spin configuration $\{\sigma^z_i\}$, $H_P$ is the total energy of the system. The initial Hamiltonian is a transverse magnetic field: $H_B = \sum_i \sigma^x_i$, where $\sigma^x_i$ is the Pauli $x$-matrix acting on qubit $i$.
The Ising Hamiltonian can be easily transformed into a quadratic unconstrained binary optimization (QUBO) problem by applying the following transformation: $w_i = \frac{\sigma_i^z+1}{2} \in \{0,1\}$. Problems submitted to DW are automatically scaled so that all $h_i$ and $J_{ij}$ values lie between $-1$ and $1$, and DW returns a set of spin values $\{\sigma_i^z=\pm 1\}$ that attempts to minimize the energy given by Eq.~\eqref{eq:ising} (a lower energy indicates better optimization). Much attention has been paid to  whether the DW devices are capable of delivering quantum speedups \cite{speedup,2014Katzgraber,Venturelli:2014nx,Hen:2015rt,Amin:2015qf}. Here we sidestep this question and instead use the DW2X as a physical device that implements quantum annealing for the purpose of solving a problem in machine learning, while focusing on performance measures other than speedup.

In order to probe the potential of a machine learning approach that is based on quantum annealing, we have used the DW2X processor to solve the simplified formulation of a biologically relevant problem: transcription factor (TF)-DNA binding (see Fig.~\ref{fig:introfig}b). TFs are a key component in the regulation of gene expression, yet the mechanisms by which they recognize their functional binding sites in a cell and thereby activate or repress transcription of target genes are incompletely understood. Nucleotide sequence, flexibility of both TFs and binding sites, the presence of cofactors, cooperativity, and chromatin accessibility are all hallmarks that affect the binding specificity of TFs in vivo \cite{Slattery:14,Shlyueva:14}. 
As a first step to gaining insight into TF binding, it is valuable to understand the intrinsic binding specificity of the TFs for DNA, which is optimally gained from in vitro data. Widely used methods to gain such an understanding and represent the DNA sequence preferences of TFs are based on position weight matrices (PWM) or PWM-like models \cite{Stormo:10}. In the simplest of these models, the binding preference of a TF for each of the four nucleotides of the DNA alphabet \{A, C, G, T\} of a sequence of length $L$ is represented as a $4\times L$ matrix. Such models implicitly treat each position in the DNA sequence as being independent, so that each element of the matrix can be thought of as the contribution of a nucleotide at the corresponding position to the overall binding affinity. Since the independence of the nucleotide positions is in many cases a valid approximation and also because of current restrictions on the size of the DW processors, in this work we have used a model consisting of single-nucleotide sequence features to show a proof of principle of the use of machine learning via quantum annealing in biology. Despite technological limitations of emerging quantum technology, we concurrently demonstrate cases in which this form of machine learning using quantum annealing outperforms classical machine learning when training with small datasets. This is among the very first successful applications of quantum hardware to a realistic, though simplified problem in computational biology. 

\section*{Results}

Experimental datasets on TF-DNA binding for a specific TF consist of $N$ sequences of fixed length $L$ and $N$ values that express a measure of the binding affinity of the chosen TF to each sequence: $\{(\vec{x}_n,y_n)\}_{n=1}^N$. In other words, the $n$th sequence is represented by the vector $\vec{x}_n = (x_{n,1},x_{n,2},\dotsc,x_{n,L})$ with $x_{n,j} \in \{$A,C,G,T$\},$ for $j = 1, \dotsc, L$, and $y_n$ is the corresponding measure of binding affinity. For instance, $\vec{x}_n$ may be ACAACTAA, with $y_n = 4.95$. {In this work we used binding from three genomic-context protein binding microarray (gcPBM) experiments, which use fluorescence intensity as a measure of binding affinity, \cite{Gordan:13} and two high-throughput systematic evolution of ligands by exponential enrichment (HT-SELEX) \cite{Jolma:10, Jolma:13, Liu:05} experiments, which report relative binding affinity. After preprocessing, the three gcPBM datasets consisted of $N\approx 1600$ sequences of $L=10$ base-pairs. The two HT-SELEX datasets consisted of $N\approx 3200$ and $1800$ sequences of length $L=12$ after preprocessing (see Methods for a brief descrption of the preprocessing procedure).} We used the following one-hot encoding to represent the sequence as a vector of binary variables: A = 1000, C = 0100, G = 0010, T = 0001, and thus transformed $\vec{x}_n$ into a feature vector $\vec{\bm{\phi}}_n \equiv (\phi_{n,1},\dotsc, \phi_{n,4L})^{\intercal}$. This encoding scheme \cite{Zhou:15} was used so that all combinations of inclusion and exclusion of the four nucleotides may be identified. Similar to previous studies \cite{Yang:14, Zhou:15, Abe:15, Yang:17} the goal of the present work is to identify patterns within the data to qualitatively assess whether the strength of a TF binding to a particular unseen sequence is above a certain threshold (classification) or to rank sequences in terms of binding affinity (ranking).

To identify conditions in which machine learning with existing quantum annealing devices may be of use for studying a simplified biological problem, we report results obtained by solving a learning protocol with six different strategies: (i) an adiabatic quantum machine learning approach formulated in Refs.~\cite{Neven1,Pudenz:2013kx} (DW), (ii) simulated annealing \cite{kirkpatrick_optimization_1983} (SA) (using the implementation given in Ref.~\cite{Isakov-SA:14}), (iii) simulated quantum annealing \cite{Santoro} (SQA), a classical algorithm that can represent the potential of a noiseless thermal quantum annealer, (iv) L$_2$ regularized multiple linear regression (MLR), (v) Lasso \cite{Tibshirani:96} and (vi) a scalable machine learning tool known as XGBoost (XGB) \cite{Chen:16}. DW, SA and SQA are probabilistic approaches. SQA is a (classical) path integral Monte Carlo method that has performed very similarly to quantum annealing and captures some of its main advantages \cite{Crosson-Harrow:2016}. MLR is a deterministic method with a closed-form solution that returns the weights that best minimize the objective function (defined below). Lasso is a method for linear regression that uses an L$_1$ norm (see description of objective function below for more details). XGB uses boosted trees and has been applied to a variety of machine learning tasks in physics, natural language processing and ad-click prediction (e.g., Ref.~\cite{chen2015higgs}).

Given a transformed feature vector $\vec{\bm{\phi}}_n$ that represents a DNA sequence, the goal of each method is to compute a predicted binding score $f(\vec{\bm{\phi}}_n)$ that best matches the actual binding score. To carry out the task, an \textit{objective function} must be optimized. The objective function consists of two parts: a training loss function and a regularization term that helps avoid overfitting. We may write the objective function as

\begin{align}
Obj(\vec{w}) = R(\vec{w})  + \Omega(\vec{w}) \label{eq:Obj},
\end{align}
where $R$ is the training loss, $\Omega$ is the regularization term, and $\vec{w}$ is the set of feature weights to be determined by the six learning algorithms: DW, SA, SQA, MLR, Lasso and XGB. The mean squared error was used as the loss function for all six methods; namely, $R(\vec{w}) = \sum_n (y_n- f_{\vec{w}}(\vec{\bm{\phi}}_n))^2$, where $y_n$ is the actual binding score of the $n$th sequence, and $f_{\vec{w}}(\vec{\bm{\phi}}_n)$ is the predicted binding score. {The regularization term was $\Omega(\vec{w}) = \lambda ||\vec{w}||_1$ for DW, SA, SQA and Lasso}, $\Omega(\vec{w}) = \lambda ||\vec{w}||_2^2$ for MLR, and $\Omega(\vec{w}) = \gamma S + \frac{1}{2}\lambda\sum_{j=1}^Sw_j^2$ for XGB, where the $\|\cdot\|_1$ norm is the number of $1$'s (Hamming weight), the $\|\cdot\|_2^2$ norm is the square of the Euclidean norm, and $S$ is the number of leaves \cite{Chen:16}. The calibration of the hyper-parameters $\lambda, \gamma$ and $S$ is discussed below. The loss function should be minimized and the regularization term generally controls model complexity by penalizing complicated models; the strength of the regularization was determined using a $100$-fold Monte Carlo cross-validation. All six methods assume a linear model for the predicted binding affinity, i.e., $f_{\vec{w}}(\vec{\bm{\phi}}_n) = \vec{w}^\intercal \vec{\bm{\phi}}_n = \sum_j w_j\phi_{n,j}$. DW, SA and SQA return binary weights and are probabilistic methods, that is, they return a distribution of weights with different energies [values of the Hamiltonian in Eq.~\eqref{eq:ising}]. In order to utilize the distribution of weights returned, while not sacrificing the discrete nature of the QUBO approach, up to twenty of the best weights were averaged (see Supplementary Material, Sec.~SI\,D for a description of how excited-state solutions were included and Fig.~S2 for an example of the decrease in the objective function).

\begin{figure}[htp]
 \includegraphics[scale=0.085]{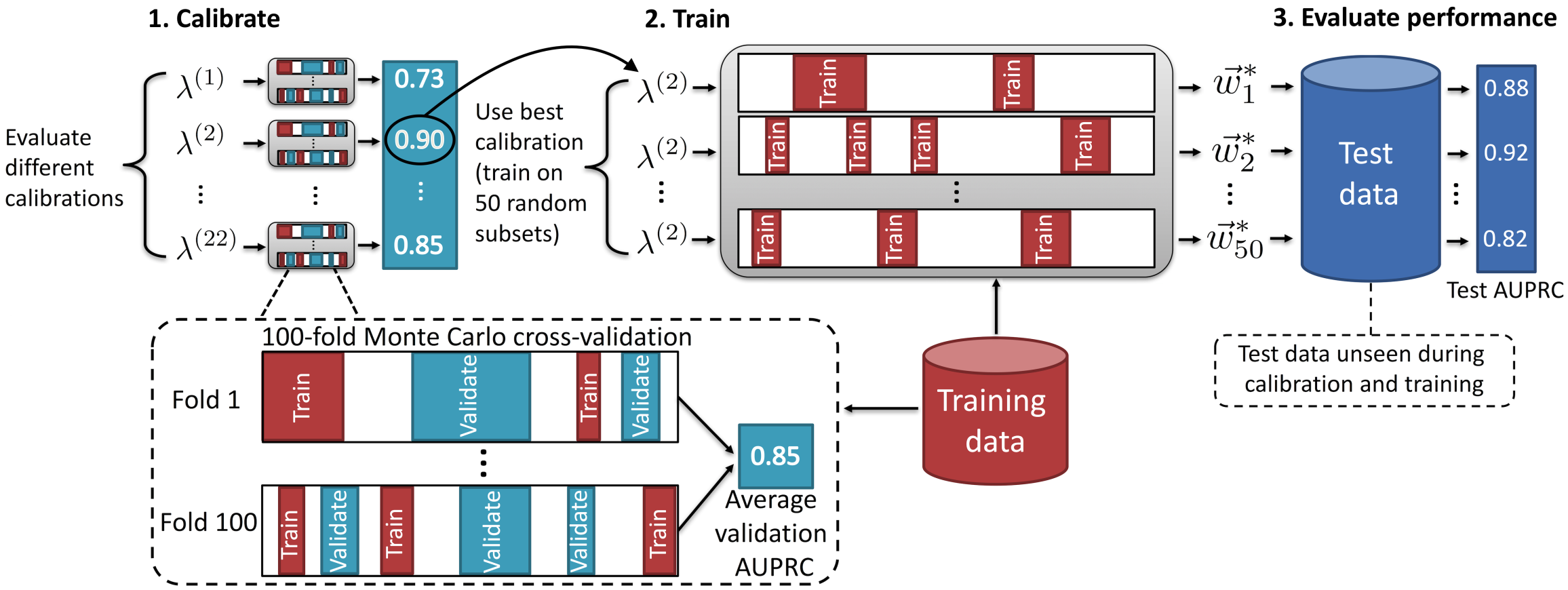}
 \caption{Schematic overview of the data handling procedure. About 10\% of the data were held out for testing ($\mathcal{D}^\textrm{TEST}$) and was unseen during the calibration and training phases. The remaining 90\% of the data were used as calibration and training ($\mathcal{D}^\textrm{TRAIN}$). For DW, SA, and MLR, in the calibration step, hyper-parameter $\lambda$ was tuned using a $100$-fold Monte Carlo cross-validation. For XGB, several other hyper-parameters were tuned (see Methods for a list). During the training step, a procedure similar to bagging (bootstrap aggregating) was used by randomly sampling a $2\%$ and $10\%$ replacement of the data $50$ times to give $50$ training instances. In the testing step, the area under the precision-recall curve (AUPRC) of the best performing weights for each of the $50$ training instances was evaluated on $\mathcal{D}^\textrm{TEST}$, which was unseen during training and calibration, to evaluate generalization of classification performance. The calibration, training, and testing procedure was identical for ranking tasks, with the exception that Kendall's $\tau$ was used as the metric of performance instead of the AUPRC.}\label{fig:DataHandling}
\end{figure}

Our computational procedure consisted of three main phases: (1) calibration of hyper-parameters, (2) training, and (3) testing (Fig.~\ref{fig:DataHandling}). About $10\%$ of the data were held out for testing during the testing phase (``test data" or $\mathcal{D}^\textrm{TEST}$); this test data was not seen during calibration and training stages. Calibration and training were carried out using the remaining $90\%$ of the data (``training data'' or $\mathcal{D}^\textrm{TRAIN}$). Due to the discrete nature of the weights returned in the QUBO approach, as well as technological limitations of the DW2X device, calibration of hyper-parameter $\lambda$ was carried out by repeatedly sampling a small number of sequences, about $2\%$ and $10\%$ of $\mathcal{D}^\textrm{TRAIN}$, corresponding to about $30$ and $150$ sequences, respectively. In particular, in the calibration phase we determined the hyper-parameters by using $100$-fold Monte Carlo (or split and shuffle) cross-validation with training splits of $2\%$ and $10\%$ of the training data, varying $\lambda$ from $2^{-3}$ to $2^{6}$. Monte Carlo cross-validation was used so that hyper-parameters would be tuned on a similar number of sequences as used in the training phase (in contrast, $n$-fold cross-validation trains on $\frac{n-1}{n}\times 100\%$ of the data). The same calibration procedure was applied to tune $\lambda$ for SA, SQA, MLR and Lasso: the resulting values of $\lambda$ are listed in the Supplementary Material, Tables S1 and S2. In order to demonstrate good performance for XGB, $\gamma$, $S$, and several additional parameters needed to be tuned (see Methods). In the training phase we used a bagging (bootstrap aggregating) procedure \cite{Breiman:96}, randomly sampling with replacement $2\%$ and $10\%$ of the training data, namely about $30$ and $150$ sequences. Each subset of about $30$ or $150$ sequences formed a training ``instance", and the mapping of a subset of data to the $h_i$ and $J_{ij}$ seen by DW and SA is given in the Methods. Each learning approach (DW, SA, SQA, MLR, Lasso and XGB) was trained on the same set of instances. To collect statistics, $50$ instances were randomly selected with replacement, for each training size. In the testing phase, the predictive power was assessed in terms of classification performance (the mean area under the precision-recall curve or AUPRC) and ranking performance (the median Kendall's $\tau$) on the test data unseen during calibration and training phases. AUPRC is a measure of classification performance that may help discern between similar algorithms when there is a high degree of class imbalance; i.e., when the data set contains many more false labels than true labels \cite{Davis:06}. Kendall's $\tau$ is a rank correlation coefficient that counts all mismatches equally \cite{Kendall:08}. Additional methodological details are given in Methods and in the Supplementary Material, Sec.~SI.

\subsection*{Performance on gcPBM Data}
\label{sec:quantComp}

To quantify the relative performance of the algorithms in capturing DNA-protein binding preferences, we first present results for high-quality gcPBM \cite{Gordan:13} data of three TFs from the basic helix-loop-helix (bHLH) family: the Mad1/Max heterodimer (`Mad'), the Max homodimer (`Max'), and the c-Myc/Max heterodimer (`Myc') \cite{Zhou:15}. bHLH proteins typically recognize and bind as dimers to the enhancer box (E-box), which is of the form CANNTG, where N denotes any of the 4 nucleotides (A, C, G, or T). Mad, Max, and Myc are part of a gene network that controls transcription in cells; a mutation of Myc has been associated with many forms of cancer \cite{Grandori:00}. For the work here, these three datasets were modified to consist of about $1600$ sequences of $10$ base pairs (bp) in length with the E-box located at the central 6 bp.

\begin{figure}[htp]
\includegraphics[scale=0.8]{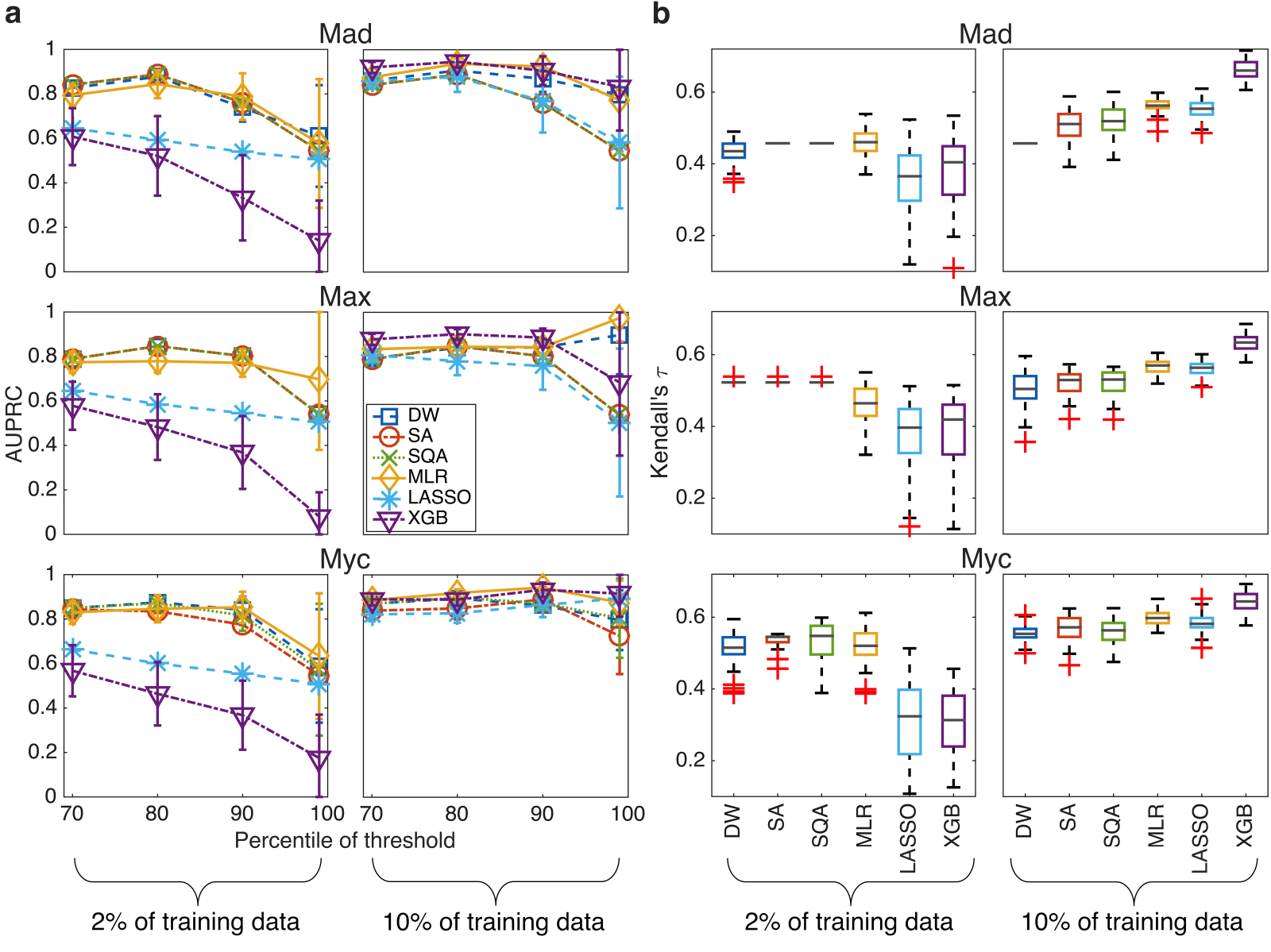}
\caption{Quantitative performance on held-out experimental test dataset of two different types of tasks for three high quality gcPBM datasets. (\textbf{a}) The mean AUPRC for Mad, Max, and Myc plotted versus threshold at certain threshold percentiles of the data, when training with 2\% of the data (left) and 10\% of the data (right). In both cases $50$ instances were randomly selected for training and performance of the 50 trained weights is evaluated on the same held-out test set. Error bars are the standard deviation. (\textbf{b}) Boxplot of Kendall's $\tau$ on held-out test dataset. Red `+' indicate outliers, gray line represents the median. The bottom and top edges of the box represent the $25$th and $75$th percentiles, respectively.}\label{fig:PBMQuantResults}
\end{figure}

In Fig.~\ref{fig:PBMQuantResults} we present the AUPRC and Kendall's $\tau$ obtained with the different algorithms when training with about $30$ (2\%) and $150$ (10\%) sequences. To compute the AUPRC, a threshold of the data was introduced: for a threshold at the $p$th percentile of the data, $p\%$ of the total number of sequences have binding affinities below the threshold and were set as negatives (``false"), and the $(1-p)\%$ of the sequences that have binding affinities above the threshold were set as positive (``true"); see Supplementary Material, Sec.~SI\,D for a more detailed explanation of the procedure to threshold the data and to generate and calculate the AUPRC. During the calibration phase, we tuned hyper-parameters with a single threshold at the $80$th percentile of the data, and during the testing phase we evaluated performance between the $70$th and the $99$th percentiles of the data. Kendall's $\tau$ was evaluated between the predicted  and measured binding affinity. A higher AUPRC indicates a better ability to correctly classify sequences that would be strongly bound by a TF, and a higher $\tau$ indicates a better ability to accurately rank the binding affinities for different sequences.

For the AUPRC, when training on instances with $2\%$ of the data (left column in Fig.~\ref{fig:PBMQuantResults}a), {DW, SA and SQA} perform very similarly, with DW slightly outperforming SA on the Myc data, and are somewhat better than MLR at the $70$th and $80$th percentiles. MLR tends to do better at the higher thresholds: this behavior could be affected by the fact that, during the calibration phase, we selected the $\lambda$ that gave the best performance at the $80$th percentile. {Lasso, which uses the same L$_1$ norm as DW, SA, and SQA, performs better than XGB but worse than the other methods}. XGB, which has been successfully applied to a growing number of learning tasks, does poorly with small training sizes. When training with $10\%$ of the data (right column in Fig.~\ref{fig:PBMQuantResults}a), the trends of relative classification performance are quite different. XGB and MLR perform very similarly, though XGB does slightly better for the Max dataset. {DW tends to perform better than SA and SQA, especially at higher thresholds. DW's mean AUPRC is normally worse than MLR and XGB's, though there is overlap between the error bars.} SA and SQA generally perform worse than the other methods, but not conspicuously so. A more thorough analysis of DW's classification performance in comparison to SA and SQA with the same problem parameters is reported in the Supplementary Material, Figs.~S3-S6 and related text in Sec.~SII\,A. {Lasso's performance is in general comparable to DW, SA, and SQA and generally seems to perform the worst with 10\% of training data.}

{For Kendall's $\tau$ (Fig.~\ref{fig:PBMQuantResults}b), Lasso and XGB's performance are the least favorable when training with $2\%$ of the data. SQA generally performs the best over the three TFs, though MLR's median $\tau$ is marginally greater than SQA's for Mad. SA's performance is very close to SQA's and DW's performance is slightly worse than the other two annealing schemes, though generally better than the typical machine learning algorithms}. With $10\%$ of the data, {DW performs the worst; SA and SQA perform very similarly, with SQA being slightly better on two of the three datasets; MLR and Lasso perform very similarly, though MLR looks slightly better; and XGB performs the best.} 

The fact that for Mad and Max with 2\% of the training data there is very little variation in Kendall's $\tau$ for SA and SQA (and to a lesser extent, DW), is a consequence of the choice of hyper-parameters. The specific values of the hyper-parameters {that gave optimal value of Kendall's $\tau$ during the calibration phase} are shown in Supplementary Material, Table S2, but we note here that the value of $\lambda$ is quite high. $\lambda$ controls the model complexity and is closely related to the bias-variance tradeoff, which states that it is impossible to simultaneously minimize errors from both bias and variance. A large value of $\lambda$ introduces a large bias \cite{Cucker:02}; consequently, for the cases where there is no or little variance, SA and SQA are essentially extracting the same pattern from all the training data. For the ranking tasks shown here with training on about $30$ sequences, this gives the best performance for SA and SQA. It may be unsurprising, however, that a large value of $\lambda$ be appropriate for small datasets; over-fitting may be a greater concern with smaller amounts of data. 

The results presented in Fig.~\ref{fig:PBMQuantResults} suggest a precise case where current quantum technology may offer slight performance advantages relative to classical computational approaches; that is, when there is only a small amount of experimental training data available (about $30$ sequences in our specific cases). In both classification and ranking tasks, DW performs comparably to SA and SQA and better than Lasso and XGB. MLR performs comparably with the annealing methods, but its error bars are much larger, indicating that its performance is less stable and more dependent on the training data. Moreover, the similarity between DW and SQA suggests that for small training sizes DW is functioning very nearly like a noiseless quantum annealer as captured by quantum Monte Carlo simulations. On a larger size of the training data DW's performance decreases relative to the classical approaches for all three TFs, though results are still competitive. The decrease in the performance of all annealing methods (DW, SA, and SQA) seems to indicate a limitation on using methods with discrete weights, which enforce simpler models. Such models may be more advantageous with a small number of training samples because they prevent overfitting. However, with larger amounts of training data, a simpler model may not adequately learn the variation within the data and hence suffer worse performance. Nevertheless, the fact that Lasso uses the same L$_1$ norm as the annealing methods (i.e., DW, SA and SQA), yet does not perform as well, indicates an advantage of such annealing methods when training with a small number of sequences. This is consistent with the finding reported in Ref. \cite{Mott:2017aa}.
 
\subsection*{Weight Logos from Feature Weights}\label{sec:consensusSequence}
\begin{figure}[htp]
\includegraphics[scale=0.8]{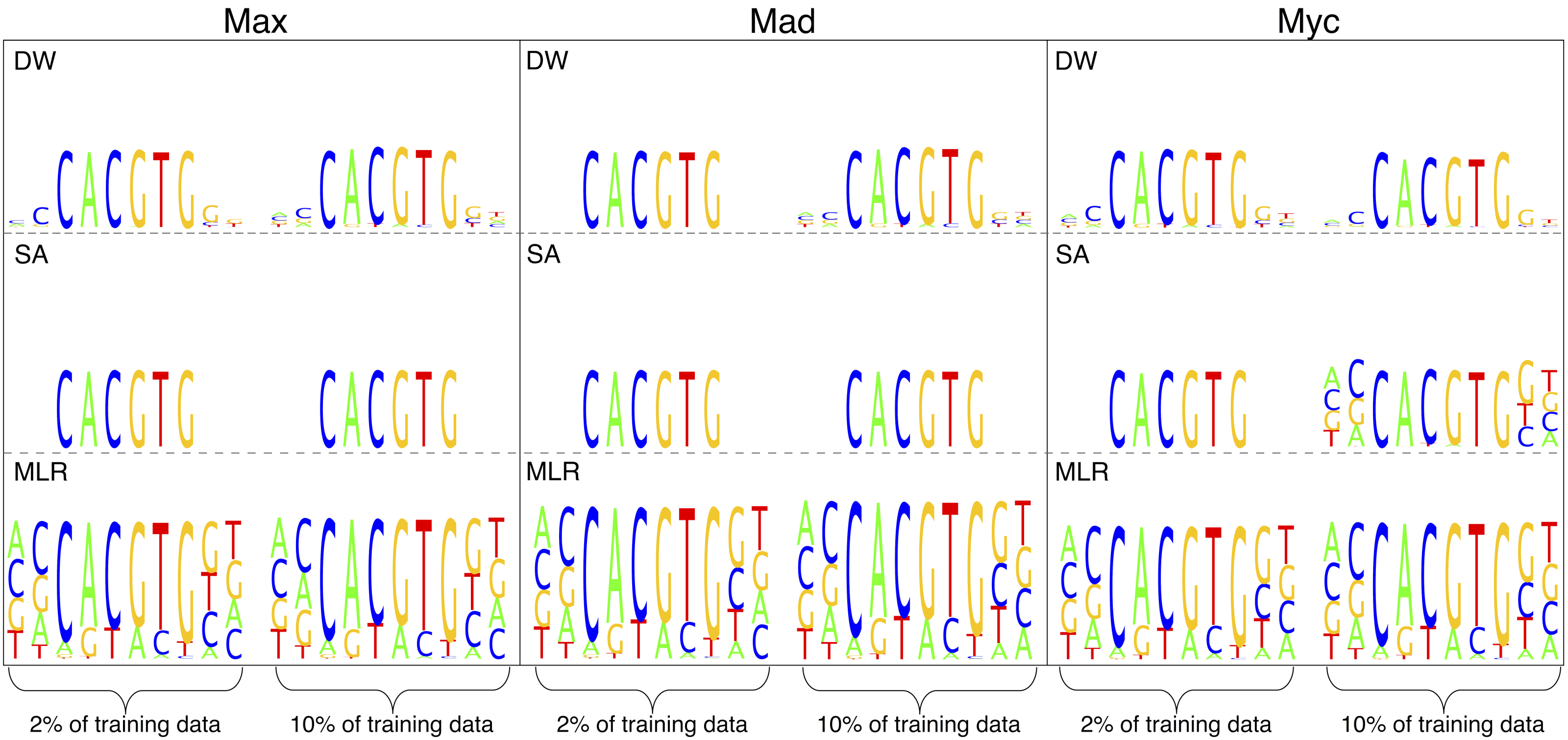}
\caption{Comparison of feature weights visualized as weight logos for DW, SA, and MLR. Weights represent the relative importance of a nucleotide at each position for the binding affinity. These weight logos were obtained using the Mad, Max, and Myc gcPBM datasets when training with the AUPRC as the objective.}\label{fig:PBMWeightLogos}
\end{figure}

Since the one-hot encoding was used with a linear model to represent DNA sequence, the weights returned by DW, SA, and MLR reflect the relative importance of the corresponding nucleotide at a position in the sequence for the binding score. The magnitude of these feature weights for DW, SA, and MLR can be visualized as a ``weight logo'' and are presented in Fig.~\ref{fig:PBMWeightLogos} for the Mad, Max, and Myc gcPBM datasets. XGB, which finds an ensemble of trees, does not assign weights to individual nucleotides and hence does not easily lend itself to visualization. Similar plots for SQA and Lasso are shown in the Supplementary Material Sec.~SII\,B and Fig.~S7. The weight logos show the contribution of nucleotides at particular positions to the strength of binding. The contribution of a nucleotide at a particular position in the sequence is represented by its height; nucleotides with the smallest weights are at the bottom and those with the largest weights are at the top. These weight logos in Fig.~\ref{fig:PBMWeightLogos} were obtained by averaging the weights from the $50$ training instances of the same number of sequences with the AUPRC as the objective. In other words, the logo represents the average of the weights that give the AUPRCs shown in Fig.~\ref{fig:PBMQuantResults}a. DW, SA, and MLR all perform very similarly and give weight logos that are in good agreement with the expected consensus sequence, CANNTG. This demonstrates that all methods are able to capture biologically relevant information. 

\subsection*{Performance on HT-SELEX Data}
\label{sec:TCF4}
HT-SELEX \cite{Jolma:10, Liu:05} is a method for investigating the relative binding affinity of a TF for a particular sequence of DNA, an in vitro technique complementary to PBM. We present results for the Max homodimer and TCF4, another member of the bHLH family with consensus sequence CANNTG, using data from HT-SELEX experiments \cite{Yang:17}. The Max dataset consisted of $3200$ sequences of $12$ bp in length, and the TCF4 dataset was modified to contain $1800$ sequences of $12$ bp in length. 

\begin{figure}[htp]
\includegraphics[scale=0.4]{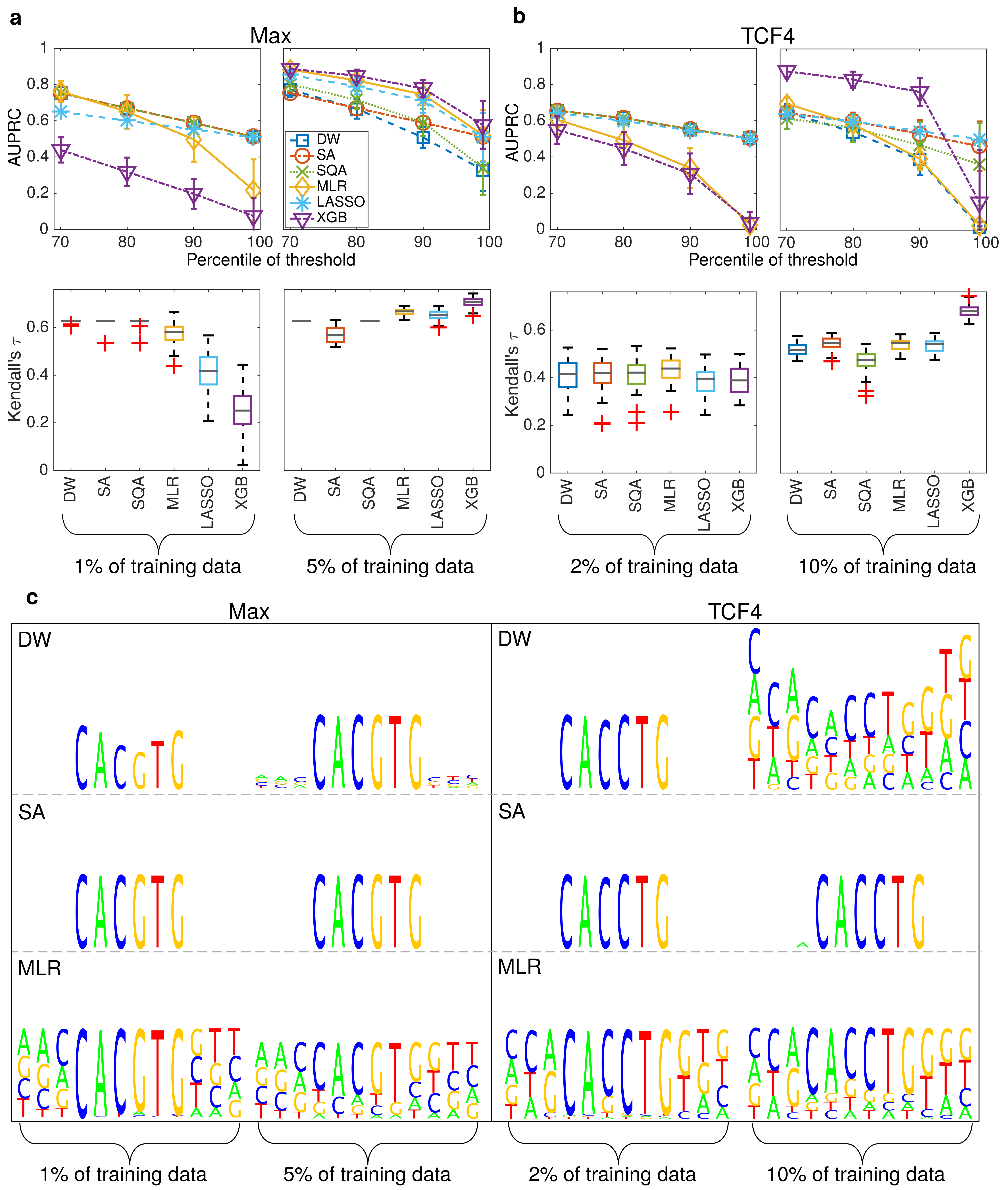}
\caption{Summary of results for HT-SELEX data. (\textbf{a}) Comparison of the AUPRC (top) and Kendall's $\tau$ (bottom) when training with 1\% of the data (left) and 5\% of the data (right) for Max. Error bars are standard deviation over 50 instances.  (\textbf{b}) Comparison of the AUPRC (top) and Kendall's $\tau$ (bottom) when training with 2\% of the data (left) and 5\% of the data (right) for TCF4. (\textbf{c}) Weight logos for Max and TCF4 from training with the AUPRC as the objective.
}
\label{fig:SelexAllResults}
\end{figure}

The {procedure for} splitting each dataset into test and training data was similar to that described earlier for the gcPBM datasets (see Fig.~\ref{fig:DataHandling}). There was no overlap between training and testing data. The quantitative results for classification and ranking performance of the six different machine learning approaches are summarized in Figs.~\ref{fig:SelexAllResults}a and \ref{fig:SelexAllResults}b, and the weight logos for DW, SA, and MLR in Fig.~\ref{fig:SelexAllResults}c. As with the gcPBM data, when training with about $30$ sequences ($1\%$ of training data for Max and $2\%$ of the training data for TCF4), DW, SA and SQA exhibit the best performance on the test dataset. MLR matches the annealing protocols with a threshold at the $70$th and $80$th percentile of the data, but does worse at the higher percentiles of the data (left column in Figs.~\ref{fig:SelexAllResults}a and \ref{fig:SelexAllResults}b). Lasso and XGB have the poorest performance. When training with about $150$ ($5\%$ of training data for Max and $10\%$ of the training data for TCF4) sequences, XGB performs very well, as on the gcPBM datasets with more training data, and MLR does well on the Max dataset but rather poorly on the TCF4 dataset; Lasso is comparable to MLR. DW's performance is worse than the best performing method (XGB), but comparable to the other methods (right column in Fig.~\ref{fig:SelexAllResults}a and ~\ref{fig:SelexAllResults}b). XGB's performance on the TCF4 dataset is much better than the other methods, except when thresholding at the $99$th percentile. 

In terms of Kendall's  $\tau$ (Fig.~\ref{fig:SelexAllResults}a and ~\ref{fig:SelexAllResults}b, bottom), all methods have similar performance when training with about $30$ sequences, with the exception of XGB which does not do as well on the Max dataset. When training with about $150$ sequences, XGB gives the best ranking performance, as it did with the gcPBM data, and the other methods all perform similarly. Finally, the weight logos in Fig.~\ref{fig:SelexAllResults}c indicate that DW, SA, and MLR capture patterns in the data that give good agreement with the expected consensus sequence. The weight logos for the Max and TCF4 HT-SELEX datasets from SQA and Lasso are reported in Supplementary Material Fig.~S8.

\section*{Discussion}
In this work we have explored the possibility of using a machine learning algorithm based on quantum annealing to solve a simplified but actual biological problem, the classification and ranking of TF-DNA binding events. This is the first application of quantum annealing to real biological data.

We have shown that DW performs comparably or slightly better than classical counterparts for classification when the training size is small, and competitively for ranking tasks.  {This trend is consistent with results on older sets of gcPBM and HT-SELEX data for various TFs, which are reported in Supplementary Material, Sec.~SIII, Figs.~S9-S14.} Moreover, these results are consistent with a similar approach for the Higgs particle classification problem \cite{Mott:2017aa}, where DW and SA both outperformed XGB with small training sizes, with a slight occasional advantage for DW over SA. This robustness across completely different application domains suggests that these findings represent real present-day advantages of annealing approaches over traditional machine learning in the setting of small-size training data. In areas of research where datasets with a small number of relevant samples may be more common, a QUBO approach such as quantum annealing realized via DW may be the algorithm of choice. On the other hand, when data is plentiful, some of the other state-of-the-art classical algorithms may be a better choice. 

We have also demonstrated that the feature weights obtained by DW reflect biological data; the weight logos for the TF-DNA binding data from gcPBM and HT-SELEX are consistent with the consensus binding site. This gives some confidence that quantum annealing is learning relevant biological patterns from the data. Yet, the approach is not without limitations. One limitation comes from the use of a single-nucleotide model to encode the DNA binding sites. In fact, we implicitly used a simple model that assumes independence between positions in the sequence. This is not always a valid approximation; higher-order ``k-mer" features or other ``shape" features that account for interdependencies between nucleotide positions may enhance model precision \cite{Mordelet:13,Zhou:13,Yang:14,Zhou:15,Yang:17}. We are limited to this simple model because of major technological constraints on the number of available qubits, which limits the number of features that can be used and thus the length of sequences that can be examined. The DW2X processor used for this study has $1098$ functional qubits, but because of a sparse connectivity between qubits, only $40$ or so features can actually be implemented on the device and in our study (see Supplementary Material, Sec.~SI\,A for more details). {Another serious limitation is the use of discrete weights. Discrete weights seem to be advantageous with a small number of training samples, as they enforce simpler models and are less prone to overfitting. However, as the amount of training data increases, these simpler models do not fare as well as some of the classical methods, which allow for greater numerical precision in the weights.}

Despite these limitations, it is encouraging to see competitive performance for the simplified problem we have studied here. Although the performance advantage from annealing-type optimizers makes it difficult to solely attribute the performance to quantumness, this work may inspire future investigations into the power of quantum annealing devices. As quantum technology continues to develop and advance, it is possible that some of the practical limitations will be addressed and the range of problems that can be explored will be expanded.  

\section*{Methods}

\subsection*{QUBO mapping of TF-DNA binding problem}
After processing the experimental datasets of $N$ sequences of fixed length $L$ and a measure of the binding affinity, we obtained the restricted datasets to which we applied six different machine learning strategies. Datasets were formulated as $\{(\vec{\bm\phi}_n,y_n)\}_{n=1}^N$, where  $\vec{\bm{\phi}}_n \equiv (\phi_{n,1},\dotsc, \phi_{n,4L})^{\intercal}$ is the transformed feature vector, and $y_n$ is the binding affinity. Solving for the simplest model is equivalent to finding a vector of binary weights $\vec{w} = (w_1,\dotsc,w_{4L})$, where $w_i \in \{0,1\}$, such that the quantity
\begin{equation}
\delta = \sum_{n=1}^N \left(y_n - \vec{w}^\intercal\vec{\bm{\phi}}_n\right)^2\label{eq:delta} 
\end{equation}
 is minimized. The problem can then be specified as finding a $\vec{w}_{\text{opt}}$ such that

\begin{align}
\vec{w}_{\text{opt}} = \argmin_w\sum_{n=1}^N \left(y_n - \vec{w}^\intercal\vec{\bm{\phi}}_n\right)^2 + \lambda||\vec{w}||_1 \label{eq:weights},
\end{align}
where $\lambda$ is a regularization (penalty) term included to prevent overfitting and $||\vec{w}||_1 = \sum_m w_m$ is the number of non-zero weights. To represent the above as an Ising problem, note that we can rewrite Eq.~\eqref{eq:weights} as follows:

\begin{align}\label{eq:isingform}
\vec{w}_{\text{opt}} &= \argmin_{\vec{w}} \sum_{n=1}^N \left(y_n - \vec{w}^\intercal\vec{\bm{\phi}}_n\right)^2 + \lambda\sum_{m=1}^{4L}w_m\\ \nonumber
& = \argmin_{\vec{w}} \sum_n \left(y_n^2 -2y_n\vec{w}^\intercal\vec{\bm{\phi}}_n + \vec{w}^\intercal\vec{\bm{\phi}}_n\vec{\bm{\phi}}_n^\intercal\vec{w}\right) + \lambda\sum_{m=1}^{4L}w_m\\ \nonumber
&= \argmin_{\vec{w}}\vec{w}^\intercal Q \vec{w} + \vec{w}^\intercal k,
\end{align}

where
\begin{align}
Q &= \sum_n \vec{\bm{\phi}}_n\vec{\bm{\phi}}_n^\intercal; \,\, Q_{i,j} = \sum_n \phi_{n,i}\phi_{n,j} \label{eq:hJ} \\
k &= \lambda\mathbf{1} -2\sum_ny_n\vec{\bm{\phi}}_n;\,\, k_i = \lambda - 2\sum_ny_n\phi_{n,i}. \nonumber
\end{align}
Constants that do not affect the optimization are dropped in the latter step. This procedure demonstrates that the problem of TF-DNA binding can be formulated as a QUBO problem, which in turn can easily be transformed into an Ising Hamiltonian of the form in Eq.~\eqref{eq:ising} and passed to D-Wave. The data normalization procedure is described in Supplementary Material, Sec.~SI\,C.

\subsection*{Technical details of algorithms}
In order to solve practical problems of interest on DW, an embedding procedure must be used (see Supplementary Material, Sec.~I\,A). Some additional preprocessing was also performed for DW and SA to ensure that all response values were feasible (see Supplementary Material, Sec.~SI\,C). 
DW, SA, SQA, MLR, Lasso and XGB were run on the same set of instances for assessment of the quantum annealer on the chosen problem. The experimental quantum processor, DW2X, was designed and built by D-Wave Systems, Inc. For each instance a total of $10000$ anneals (``runs") were collected from the processor, run with an annealing time of $20\mu$s. SA and SQA are classical analogues of quantum annealing that perform annealing on a classical and path integral Monte Carlo simulation of the Ising spin glass, respectively. SA and SQA were run with $10000$ sweeps (each sweep is an update of all spins) per repetition (or ``anneals") with an initial inverse temperature of $0.1$ and a final inverse temperature of $3$, for a total of $10000$ repetitions. The SA code was adapted from Ref. \cite{Isakov-SA:14}, and an in-house version of SQA was used. MLR is a widely used technique to minimize the loss function shown in Eq.~\eqref{eq:weights}, with the convex penalty term $\lambda ||\vec{w}||_2^2$ instead of the linear penalty term. Lasso has the linear penalty term \cite{Tibshirani:96}; XGB uses boosteed trees \cite{Chen:16}. The weights $\vec{w}$ returned by MLR, Lasso and XGB are real-valued, whereas the weights returned by DW, SA and SQA (which solve a QUBO/Ising problem) are binary. In addition, DW, SA and SQA are probabilistic, meaning that a distribution of weights with different energies [the value of $H_P$ in Eq.~\eqref{eq:ising}] are returned. Up to $20$ of the lowest energy weights were included for both DW, SA and SQA (see Supplementary Material, Sec.~SI\,D for more details). The lower the energy, the better the particular solution is at minimizing Eq.~\eqref{eq:weights}. In contrast, MLR, Lasso and XGB are deterministic and return a single solution. 

In the calibration phase, only one hyper-parameter, $\lambda$ was tuned for DW, SA, SQA, MLR and Lasso. All five methods were tuned separately for both classification and ranking tasks, resulting in different optimal $\lambda$ for each method (see Supplementary Material Tables S2 and S3 for final values of $\lambda$). With an older dataset we varied both the number of sweeps for SA and the value of $\lambda$ but results were not significantly different; hence, here we only vary $\lambda$ for SA. SA also has various other parameters that are related to the algorithm itself, including number of runs, initial and final temperature, and the cooling schedule, all of which affect the optimization performance. These parameters were not tuned. Similar additional parameters for DW, including annealing time and number of runs, were also not tuned. XGB's performance depends on several hyper-parameters, and more careful tuning was necessary in order to give competitive performance. XGB parameters \cite{Chen:16} that were considered include $\gamma$, the max\_depth, and min\_child\_weight (all of which control model complexity), subsample, colsample\_bytree, (which add randomness to make training robust to noise), as well as learning rate, eta. Rather than doing a full grid search over all these parameters, parameters were tuned sequentially; i.e., one value of $\eta$ was fixed, then the best value of max\_depth and min\_child\_weight were found. The optimal $\gamma$ for those values was then found; and finally subsample and colsample\_bytree tuned. $\eta$ was then varied and the process repeated. $\eta$ was varied from $0.05$ to $0.3$, max\_depth from $3$ to $20$, min\_child\_weight from $1$ to $20$, $\gamma$ from $0$ to $1$, and subsample and colsample\_bytree both from $0.6$ to $1$.

In the testing phase, we evaluated performance based on two metrics:  the AUPRC for classification performance and Kendall's $\tau$ for ranking performance. For the AUPRC, we reported mean values with standard deviations as error bars, whereas for Kendall's $\tau$ the median value was presented. 

\subsection*{Data processing and availability}
\label{sec:DataandPrecision}
Original probes for the gcPBM \cite{Gordan:13} data contained $16000-18000$ sequences of 36 bp in length with the fluorescence intensity as a measure of binding affinity. The same data is used in \cite{Zhou:15} and may be downloaded from GEO (https://www.ncbi.nlm.nih.gov/geo/) under accession number GSE59845. Because of current limitations of the architecture of the DW device that limit the number of features that may be used, the data was truncated to the central $10$ bp. For each sequence of $10$ bp, we calculated its average gcPBM signal. In other words, all sequences in the datasets were unique. The final Mad, Max, and Myc datasets consisted of $1655$, $1642$, and $1584$ sequences, respectively, of length $10$ bp, and the logarithm base $2$ with fluorescence intensities was used. The HT-SELEX data came from mammalian TFs \cite{Jolma:13} that was re-sequenced with on average $10$-fold increase in sequencing depth \cite{Yang:17}. The sequencing data is available at the European Nucleotide Archive (ENA - https://www.ebi.ac.uk/ena; study identifier PRJEB14744) and was pre-processed following the protocol in \cite{Yang:17}. After this first step of pre-processing, the Max and TCF4 datasets consisted of $3209$ and $15556$ sequences of length $12$ bp and $14$ bp, respectively. The Max dataset did not require further truncation, but one bp on the left and right flanks were trimmed for the TCF4 dataset, giving a modified dataset of $1826$ sequences of length $12$ bp. As with the gcPBM data, the average relative affinity was averaged for each truncated sequence. 

\section*{Acknowledgments} The authors thank Lin Yang for assistance with pre-processing the gcPBM and HT-SELEX data. The authors would also like to thank Tameem Albash for providing the SQA code. This work was supported by the USC Women in Science and Engineering Program )to R.D.F), the National Institutes of Health grants R01GM106056 and U01GM103804 (to R.R.), ARO grant number W911NF-12-1-0523 and NSF grant number INSPIRE-1551064 (to D.L.). R.R. is an Alfred P. Sloan Research Fellow. D.L. is a Guggenheim Foundation Fellow and a Caltech Moore Scholar in Physics. This research is based upon work (partially) supported by the Office of the Director of National Intelligence (ODNI), Intelligence Advanced Research Projects Activity (IARPA), via the U.S. Army Research Office contract W911NF-17-C-0050. The views and conclusions contained herein are those of the authors and should not be interpreted as necessarily representing the official policies or endorsements, either expressed orimplied, of the ODNI, IARPA, or the U.S. Government. The U.S. Governmentis authorized to reproduce and distribute reprints for Governmental purposes notwithstanding any copyright annotation thereon.

\section*{Competing Interests}
The authors declare no competing financial interests.

\section*{Author contributions}
R.L, R.D.F, R.R. and D.L designed and conceived the study. R.L. implemented and executed DW, SA, SQA, MLR, Lasso and XGB, and analyzed the results. R.R. and D.L directed the study. R.L, R.D.F, R.R., and D.L wrote the manuscript.


\clearpage 

\section*{Figure Legends}

\end{document}